\documentclass[aps,prd,twocolumn,superscriptaddress,nofootinbib,longbibliography]{revtex4-2}
\usepackage{bm, amsmath, amsfonts, amssymb, braket}
\usepackage{subfigure}
\usepackage[dvipsnames]{xcolor}
\usepackage{ulem,soul,changes}
\usepackage{dcolumn} 
\usepackage[colorlinks]{hyperref}
\AtBeginDocument{
	\hypersetup{
		urlcolor=black,
		citecolor=green,
		linkcolor=blue,
	}%
}
\usepackage{romannum}
\usepackage{cases,float}
\usepackage[toc,title,page]{appendix}

\usepackage{verbatim}
\usepackage{amstext}
\usepackage{array}
\usepackage{multirow}
\usepackage{physics}
\usepackage{geometry}
\usepackage{enumitem}
\usepackage{indentfirst}

\AtBeginDocument{\pagenumbering{arabic}}

\begin{document}
	
\title{Gravito-diamagnetic forces for mass independent large spatial superpositions}
	\newcommand{\affone}{Centre for Quantum Computation and Communication Technology, School of Mathematics and Physics, University of Queensland, Brisbane, Queensland 4072, Australia}
	\newcommand{\afftwo}{Department of Physics and Astronomy, University College London, Gower Street, WC1E 6BT London, United Kingdom.}
	\newcommand{\affthree}{Van Swinderen Institute, University of Groningen, 9747 AG Groningen, The Netherlands.}

	\author{Run Zhou}
	\affiliation{\affthree}
	\author{Ryan J. Marshman}
	\affiliation{\affone}
	\author{Sougato Bose}
	\affiliation{\afftwo}
	\author{Anupam Mazumdar}
	\affiliation{\affthree}
	
	\date{\today}

\begin{abstract}
Creating a massive spatial quantum superposition, such as the Schr\"odinger cat state, where the mass and the superposition size within the range  $10^{-19}-10^{-14}$ kg and $\Delta x \sim 10~{\rm nm}-100~\mu {\rm m}$, is a challenging task. The methods employed so far rely either on wavepacket expansion or on a quantum ancilla, e.g. single spin dependent forces, which scale inversely with mass. In this paper, we present a novel approach that combines gravitational acceleration and diamagnetic repulsion to generate a large spatial superposition in a relatively short time. After first creating a modest initial spatial superposition of $1~\mu {\rm m}$, achieved through techniques such as the Stern-Gerlach (SG) apparatus, we will show that we can achieve an $\sim 10^{2}-10^{3}$ fold improvement to the spatial superposition size ($1~{\rm \mu m}\rightarrow 980~\mu {\rm m}$) between the wave packets in less than $0.02$~s by using the Earth's gravitational acceleration and then the diamagnetic repulsive scattering of the nanocrystal, neither of which depend on the object mass. Finally, the wave packet trajectories can be closed so that spatial interference fringes can be observed. Our findings highlight the potential of combining gravitational acceleration and diamagnetic repulsion to create and manipulate large spatial superpositions, offering new insights into creating macroscopic quantum superpositions.
\end{abstract}
\maketitle	

\section{Introduction} 

Creating large spatial superposition states on a macroscopic scale represents a cutting-edge frontier in contemporary quantum research, intersecting theoretical exploration with experimental ingenuity. This pursuit holds significant promise for testing the foundational principles of quantum mechanics in the presence of gravity \cite{penrose1996gravity,Diosi:1988tf,pearle1989collapse,bassi2013models, nimmrichter2013macroscopicity}, investigating the equivalence principle \cite{Bose:2022czr,Chakraborty:2023kel}, placing bounds on decoherence mechanisms \cite{PhysRevA.84.052121,romero2011large,Romero_Isart_2010,Tilly:2021qef,Schut:2021svd,Rijavec:2020qxd,Fragolino:2023agd}, and exploring applications in quantum sensors \cite{Toros:2020dbf,Marshman:2018upe,Wu:2022rdv}, the detection of gravitational waves \cite{Marshman:2018upe}, and the probing of a potential fifth force \cite{Barker:2022mdz}.

The synthesis of two macroscopic matter-wave interferometers, relying on massive spatial superposition states, offers a potential avenue for laboratory testing of the quantum nature of gravity \cite{bose2017spin,ICTS,marletto2017gravitationally}. For a comprehensive understanding of the theoretical underpinnings, refer to \cite{Marshman:2018upe,danielson2022gravitationally,bose2022entanglement,carney2018tabletop,Carney_2019,Belenchia:2018szb,Elahi:2023ozf,Vinckers:2023grv,Martin-Martinez:2022uio,Fragkos:2022tbm}. It is noteworthy that the creation of a substantial Gaussian state, designed for probing the quantum interactions with a photon, is feasible \cite{Biswas:2022qto}. However, in this study, we {\it solely} focus on creating non-Gaussian quantum state.

Despite routine observations of superpositions in microscopic particles like electrons and atoms, generating superpositions in truly macroscopic objects remains a formidable challenge. To date, the heaviest masses placed in a superposition of spatially distinct states are macromolecules with a mass on the order of $10^{-23}$ kg \cite{Arndt:1999kyb, gerlich2011quantum, FeinEtAl2019}.

To unlock new opportunities for quantum-enhanced applications and gain insights into the quantum-to-classical transition, various physical schemes have been proposed to achieve superposition sizes ranging from $10~\text{nm}$ to $1~\mu\text{m}$ for large masses ($m\sim 10^{-19}-10^{-17}$ kg) \cite{Bose:1997mf,Sekatski_2014,Romero_Isart_2010,romero2011large,Hogan,wan2016free,scala2013matter,yin2013large,clarke2018growingPublished,Wood:2022htp,pedernales2020c,kaltenbaek2016macroscopic,pino2018chip,romero2017coherent,kaltenbaek2012MAQRO,arndt:2014testing}. More ambitious proposals aim at matter wave interferometers with masses up to $10^{-15}$ kg delocalized from $1~\mu\text{m}$ to $100~\mu\text{m}$ with a coherence time of around 1 s \cite{Marshman:2021wyk,zhou2022}. However, these schemes face diminishing effectiveness as mass increases due to a single spin-dependent force.

Addressing this challenge, recent work has introduced a mass-independent scheme leveraging diamagnetic repulsion to enhance spatial superposition from an initial size of $1~\mu\text{m}$ \cite{Zhou:2022jug}. Nevertheless, two key challenges impede the realization of this mass-independent approach. The first challenge involves establishing an initial spatial superposition at approximately $1~\mu\text{m}$, while the second revolves around imparting the superposition with an initial velocity, crucial for achieving larger superposition sizes within a shorter runtime.

This current study focuses on addressing the latter challenge. We assume that the initial spatial superposition splitting of ${\cal O}(1)~\mu \text{m}$ has already been established through other known mechanisms \cite{wan2016free,Pedernales_2020,Marshman:2021wyk,zhou2022,Wood:2022htp}. We discuss one of these mechanisms, specifically the utilization of a linear magnetic field combined with nitrogen-vacancy (NV) centre spin to create the initial spatial separation, see Appendix \ref{initial_condition}. Unlike previous methods that allowed the entire apparatus to free-fall to circumvent gravity \cite{Marshman:2021wyk,zhou2022}, our approach effectively utilizes gravity, offering a comprehensive, mass-independent solution for enhancing spatial superposition. This gravity-inclusive approach provides two significant advantages: it simplifies experimental design by eliminating the need to drop the apparatus within a tower, conserving space and reducing design complexity, and it offers a mass-independent mechanism for generating substantial initial velocities within a short time frame—a critical factor in achieving large spatial superposition states.

Accelerating objects becomes increasingly challenging as the mass grows, and conventional methods, such as laser pulses used for accelerating atoms \cite{kovachy2015}, are unsuitable for heavier nanoparticles like nano-particles due to heat generation and disruption of the superposition \cite{gonzalez2021}.

This paper aims to illustrate how a mass-independent scheme of diamagnetic repulsion, incorporating Earth's gravitational acceleration, can achieve a substantial enhancement of superposition size, reaching ${\cal O}(980)\,\mu {\rm m}$ in $0.02$ s. The proposed timescale closely aligns with existing experimental conditions, considering that the longest spin coherence times achieved in the laboratory are on the order of ${\cal O}(1)$ ms for nano-particles with a mass of approximately $10^{-15}$ kg \cite{TrusheimEtAl2014, Wood:2021jqe}. This coherence time can be extended further by employing purer nano-particles \cite{Frangeskou:2016mqs} and lowering temperatures \cite{Bar-Gill2013}. While nano-particles are the focus of this study, our approach theoretically applies to any substance with diamagnetic properties. The study begins with a theoretical analysis of nanoparticle motion in gravitational and magnetic fields, followed by numerical verification of our calculations.

\section{Mass-independent acceleration under gravity}

 The Hamiltonian of a diamagnetic material like diamond crystal in the presence of an external magnetic field is given by \cite{wan2016free,Pedernales_2020,Marshman:2021wyk,Wood:2022htp}: 
\begin{equation}\label{Hamitonian}
	H=\frac{\boldsymbol{p}^{2}}{2 m} -\frac{\chi_{\rho} m}{2 \mu_{0}} \vb{B}^{2} + mgz\vb*{e_{z}}.
\end{equation}	
In Eq.(\ref{Hamitonian}), the first term represents the kinetic energy of the nano-particle, where $\boldsymbol{p}$ is the momentum and $m$ is the mass of the nano-particle~\footnote{Note that the fundamental process we discuss is not exclusive to diamond crystals. Although we employ properties of diamonds for our numerical calculations, the scope of this mechanism extends further. The calculations we present are applicable to any type of diamagnetic crystal.}. The second term signifies the magnetic energy of a diamagnetic material (nano-particle) in a magnetic field, with $\chi_{\rho}=-6.2\times10^{-9}$ ${\rm m^{3}/kg}$ as the mass susceptibility and $\mu_{0}$ as the vacuum permeability. The final term denotes the gravitational potential energy, where $g\approx9.8$ ${\rm m/s^{2}}$ is the gravitational acceleration, and $\vb*{e_{z}}$ is the unit vector along the positive $z$ axis.

The potential energy experienced by the nano-particle, according to the Hamiltonian Eq.(\ref{Hamitonian}), is expressed as:
\begin{equation}\label{Potential}
	U=- \frac{\chi_{\rho} m}{2 \mu_{0}} \vb{B}^{2} + mgz\vb*{e_{z}}.
\end{equation}	
To facilitate the separation of wave packets through diamagnetic repulsion, we consider the central magnetic field generated by a current-carrying wire, given by:
\begin{equation}\label{MagneticField}
	\vb{B}=\frac{ \mu_{0} \vb*{I}\cross \vb*{e_{r}}}{2\pi r}.
\end{equation}	
Here, $\vb*{I}$ denotes the current carried by a straight wire, and $r$ is the radial distance from a point in space to the center of the wire. $\vb*{e_{r}}$ represents the unit vector in the radial direction.

Combining Eq.(\ref{Potential}) and Eq.(\ref{MagneticField}), the acceleration of the nano-particle is derived as:
\begin{equation} \label{Acceleration}
	\vb*{a_{dia}}=-\frac{1}{m}\grad{U}=\alpha\frac{I^{2}}{r^{3}}\vb*{e_{r}} - g\vb*{e_{z}}.
\end{equation}
Here, $\alpha=-\chi_{\rho} \mu_{0}/4\pi^{2}$ is defined. Notably, in Eq.(\ref{Acceleration}), the acceleration is found to be independent of the mass. This distinctive property offers an exceptional opportunity to achieve a significant superposition size for a massive quantum object. Assuming the straight wire is perpendicular to the $x-z$ plane and considering the scenario where the nano-particle starts to fall from rest, the motion of the nano-particle will be confined to the $x-z$ plane.

The physical picture presented here differs from the configuration described in our earlier work \cite{Zhou:2022jug}. That previous paper aspired to create a superposition where the Earth’s gravitational potential is zero, a situation that could potentially arise if the superposition is created in a diamagnetic trap \cite{Schut:2023hsy}. However, the dynamics differ if the goal is to create a macroscopic superposition under Earth’s gravity, as explained below.


\section{Scattering processes}\label{DiamagneticRepulsion}
\begin{figure}[h]
	\includegraphics[width=0.8\linewidth]{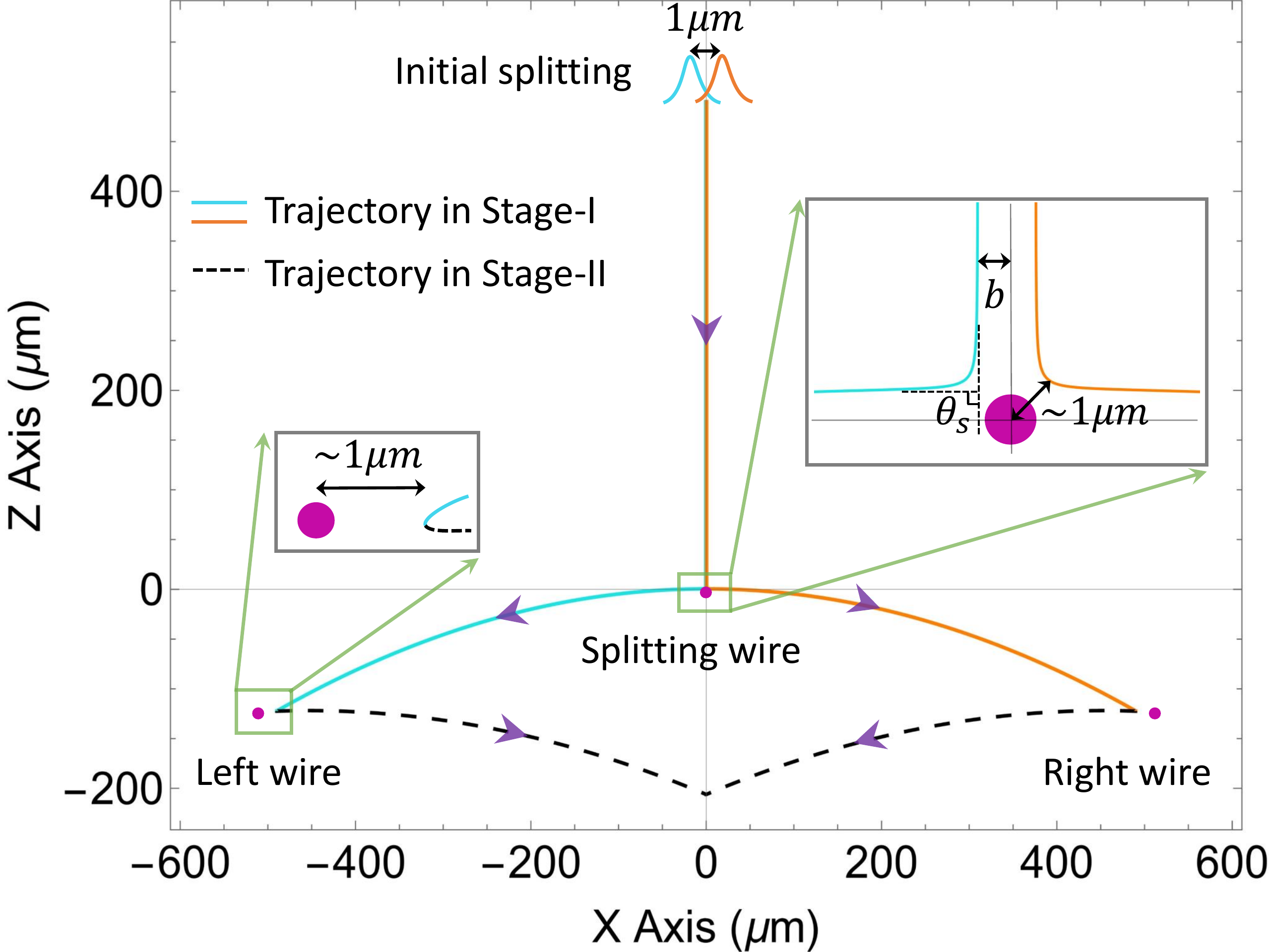}
	
	\caption{\footnotesize {Experimental scheme and numerical results for creating mass independent spatial superposition by solving Eq.(\ref{Acceleration}). The two wave packets with an initial spatial splitting of $\Delta x_{0}$ start falling, {\it under Earth's gravity acting along the $-z$ direction}, from the rest and enter the magnetic field generated by the current-carrying wires. The purple points represent straight wires perpendicular to the $x-z$ plane.  The wire at the origin is called the splitting wire located at $(x=0, z=0)$. The wire on the left is called the {\it left wire} and the wire on the right is called the {\it right wire}. The blue and orange solid lines represent the wave packet trajectory in stage-$\text{\Romannum{1}}$ and the black dashed lines represent the wave packet trajectory in stage-$\text{\Romannum{2}}$. The purple arrows indicate the direction of motion of the wave packets. The sign $b$ is the impact parameter and the initial separation $\Delta x_{0}=2b=1~{\rm \mu m}$. The sign $\theta_{s}$ is the scattering angle. The total motion time is around 0.0195 s.}} \label{GravityAccelerationNumaricalResult}
\end{figure}

The experimental setup comprises two distinct stages, as illustrated in Fig.\ref{GravityAccelerationNumaricalResult}. In Stage-\Romannum{1}, our primary objective is to maximize spatial superposition. Stage-\Romannum{2}, on the other hand, focuses on converging the trajectories of the wave packets.

Stage-\Romannum{1} is further subdivided into two components. The first component exploits the mass-independent acceleration due to gravity, allowing us to control the nano-particle's velocity by adjusting its initial height. The second component leverages the mass-independent diamagnetic acceleration, as expressed by the first term on the right side of Eq.(\ref{Acceleration}), to modify the velocity vector of the nano-particle. In scenarios where the initial positions of the two superimposed wave packets symmetrically flank the wire, they exhibit opposite velocity changes, facilitating significant spatial separation, as illustrated in Fig.\ref{GravityAccelerationNumaricalResult}.

In Stage-\Romannum{2}, these two wave packets undergo scattering and velocity redirection through a pair of symmetrical wires, one on the left and one on the right, ultimately closing their trajectories.

To support our claims, we provide a detailed analytical exposition of the nano-particle's acceleration by the gravitational field and its elastic scattering by the magnetic field generated by the current-carrying wire. Additionally, we perform numerical simulations to validate our analytical findings.

\section{Analytical treatment}
We assume that the nano-particle is accelerated by gravity from the rest at a distance from the wire and the effect of diamagnetic repulsion on the nano-particle can be ignored at the initial stages of the free-fall. We set the distance $z_{0}$ at which the nano-particle is accelerated due to gravity. The initial velocity of the nano-particle derived by the gravitational acceleration is given by:
\begin{equation} \label{IncidentVelocity}
	\boldsymbol{v_{in}}=\sqrt{2z_{0}g}\vb*{e_{z}}.
\end{equation}
The time for the nano-particle to fall from its initial position to the splitting wire is:
\begin{equation} \label{Time1}
	t_{1}=\sqrt{\frac{2z_{0}}{g}}.
\end{equation}
We now consider the process of a nano-particle incident at a velocity $\boldsymbol{v_{in}}$, and then scattered by the magnetic field generated by the current-carrying wire. This scattering process can be solved analytically, see \cite{Zhou:2022jug}. The scattering angle of the nano-particle scattered by the central magnetic field is given by \cite{Zhou:2022jug}:
\begin{equation}\label{ScatteringAngle}
	\theta_{s}=\qty(1-\frac{1}{\sqrt{k}})\pi,
\end{equation}
where
\begin{equation}\label{Frequency}
	k=1+\alpha\frac{I^{2}}{v_{in}^{2}b^{2}}.
\end{equation}
Here $b$ is the impact parameter. The geometric picture of $\theta_{s}$ and $b$ is shown in Fig.\ref{GravityAccelerationNumaricalResult}. In order to obtain the maximum superposition size in the shortest time, we set the scattering angle for the first time $\theta_{s1}=\pi/2$. We assume that the nano-particle is scattered for the second time at coordinates $(\pm x_{spl}, z_{1})$ with the left (right) wire, and that the velocity direction after scattering is parallel to the $x$-axis. If $x_{spl}=z_{0}$, then the scattering angle for the second time $\theta_{s2}\approx3\pi/4$. The time for the nano-particle to travel from the splitting wire to the left (right) wire is:
\begin{equation}\label{Time2}
		t_{2}=\frac{x_{spl}}{v_{x}}.
\end{equation}
$v_{x}$ is the component of the velocity of the nano-particle along the $x$-axis. Since the scattering angle is $\pi/2$, therefore $v_{x}=v_{in}$. And then the time for the nano-particle to travel from the left (right) wire to the $z$-axis is:
\begin{equation}\label{Time3}
	t_{3}=\frac{x_{spl}}{\sqrt{v^{2}_{x}+g^{2}t^{2}_{2}}}.
\end{equation}
Combining Eq.(\ref{IncidentVelocity}), (\ref{Time1}), (\ref{Time2}) and (\ref{Time3}) gives a total evolution time of
\begin{align}\label{TimeTotal}
	t_{tot} &= t_{1}+t_{2}+t_{3},\nonumber\\
	        &= \sqrt{\frac{2z_{0}}{g}}\qty(1+\frac{x_{spl}}{2 z_{0}}+\frac{x_{spl}}{\sqrt{4 z_{0}^{2}+x^{2}_{spl}}}).
\end{align}
It can be seen from Eq.(\ref{TimeTotal}) that when $z_{0}$ is fixed, the smaller the $x_{spl}$, the shorter the evolution time $t_{tot}$. We have ignored the deceleration and acceleration of the nano-particle as it approaches the wire in our calculations. This is because the diamagnetic repulsion is inversely proportional to the third power of the distance and only dominates when the nano-particle is very close to the wire ($\sim 10\,{\rm \mu m}$).

\subsection{Numerical results}

We use the equation of motion, Eq.(\ref{Acceleration}), to numerically solve for the trajectory of the wave packet. The numerical results are shown in Fig.\ref{GravityAccelerationNumaricalResult}. We set the initial separation between wave packets $\Delta x_{0}=1$ $\mu \text{m}$ ($b=0.5$ $\mu \text{m}$) and the initial coordinates of the classical positions of the two wave packets to be $(\pm 0.5, 490)$ $\mu \text{m}$. The coordinate of the splitting wire is $(0,0)$ and the current through it, which is determined by Eq.(\ref{ScatteringAngle}) with $\theta_{s1}=\pi/2$, is 6.04138 A. The coordinates of the left and right wires are $(\pm 491,-122.6)$ $\mu \text{m}$. We adjust the currents in the left and right wires so that the velocity direction of the wave packet is approximately parallel to the $x$-axis after the second scattering. The current through the left and right wires is $10$ A. All three wires are switched on during the experiment. 

It is important to note that fluctuations in the current can affect the accuracy of the wave packet position (expectation value) and thus the spatial interference fringe. We analysed the effect of current fluctuations on the wave packet position and the limits imposed on current fluctuations in order to produce spatial interference in Appendix \ref{current_fluctuation}.


With the parameters we have set, the total dynamical time agrees well between analytical and numerical results (0.0194742 s vs. 0.0194958 s). In the process of separating and then recombining the two wave packets, the maximum superposition size reaches about $980$ $\mu \text{m}$ and the initial separation between the wave packets is amplified by a factor of about 1000 in 0.02 s.

The closest distance of the wave packet trajectory to the splitting wire is $1.00081~\mu \text{m}$, and the closest distance of the wavepacket to the left (right) wire is $1.32289~\mu \text{m}$. If this minimum distance is considered (as an upper limit) to be the maximum radius of the wire, then the current density is $\sim1.9~{\rm A/\mu m^{2}}$ for the splitting wire, and $\sim 1.8~{\rm A/\mu m^{2}}$ for the left (right)  wire, which is {\it currently} achievable in a laboratory with carbon nanotubes and graphene \cite{Yao2000,Wei2001,Murali2009}. The current density through the wire, $\rho_{current}$, the incident velocity of the nano-particle, $v_{in}$, and the impact parameter, $b$, satisfy the relation \cite{Zhou:2022jug}
\begin{equation}\label{CurrentDensity}
	\rho_{current}=\frac{I}{\pi d^{2}}=\frac{1}{b}\frac{Cv_{in}^{2}}{\pi(v_{in}^{2}+\alpha C^{2})},
\end{equation}
where $C=I/b$ and $d$ are the closest distance of the wave packet trajectory to the centre of the wire. When the scattering angle is $\pi/2$, combining Eq.(\ref{ScatteringAngle}) and (\ref{CurrentDensity}) and then we have
\begin{equation}\label{CurrentDensity2}
	\rho_{current}=\frac{1}{4 b\pi}\sqrt{\frac{3}{\alpha}}v_{in}.
\end{equation}
From Eq.(\ref{CurrentDensity2}) it can be seen that the current density is linear with the incident velocity. Additionally, the current density is also inversely proportional to the impact parameter $b$. This means that for a smaller initial separation ($\Delta x_{0}=2b$), we need a larger current density to achieve the same superposition size.

\section{Conclusion and discussion}

In this paper, we have achieved $\sim 100-1000$ times increment in the spatial superposition, from $1~\mu {\rm m}\rightarrow 980~ \mu {\rm m}$ between the wave packets in $0.02$~s by using gravitational acceleration and the repulsive, diamagnetic scattering off the nano-particle. There are three distinct advantages to this scheme. 
(1) The first is that the process of enhancing the spatial superposition is  {\it mass independent}. In the ideal situation, when all the known Standard Model interactions are under control along with all the known sources of the decoherence, we can use this mass-independent scheme to increase the spatial superposition between the wave packets ${\cal O}(10^{3})~\mu {\rm  m}$, or even higher, provided that we can create an initial spatial superposition, even one as small as $1\,\mu {\rm m}$. This scheme solves some of the outstanding challenges of creating large spatial superposition, either using the wave packet expansions \cite{Arndt:1999kyb,gerlich2011quantum,romero2011large,kaltenbaek2016macroscopic,pino2018chip,romero2017coherent,kaltenbaek2012MAQRO,arndt:2014testing} or spin-dependent forces \cite{Margalit2021,Marshman:2021wyk,zhou2022}. In all the previous cases the efficacy is reduced by any increase in the mass. 
(2) The second advantage is that the whole process takes a shorter time (around $0.02$ s) compared to previous schemes for creating large spatial superposition \cite{Marshman:2021wyk,Wood:2022htp,zhou2022}. The shorter time in which one experimental run is performed will reduce the time during which the environment can act to decohere the system. This will also improve the total run-time of the experiment or conversely increase the number of experimental runs performed which is essential for witnessing the entanglement induced by the quantum nature of gravity~\cite{Tilly:2021qef,Schut:2021svd}.
(3) The third advantage is that the experimental apparatus (wires) is fixed. Compared to the previous scheme of creating spatial superposition \cite{Marshman:2021wyk,zhou2022,Zhou:2022jug}, where the experimental apparatus is free falling in the gravitational field, this is easier to achieve in the laboratory. It is also possible to create spatial superposition in optical or magnetic levitation systems \cite{Delic:2020ndp,hsu2016a}, but the levitation system itself limits the superposition size that can be achieved.

It is important to note that although the process of enhancing the superposition size is mass independent, there are three factors that limit the mass of the nano-particle. Firstly, an initial spatial separation between wave packets is required. If a spin-dependent force is employed for this purpose, the process of creating the initial spatial separation is still mass dependent \cite{Marshman:2021wyk,zhou2022}. Secondly, the distance between the nano-particle and the wire serves as a limiting factor for the nano-particle's mass. Assuming the maximum radius of the nano-particle to be half the minimum distance between the nano-particle and the wire ($\sim0.5{\rm \mu m}$), this results in a maximum nano-particle mass of approximately $10^{-15}$ kg. Thirdly, the stability of the current and the spatial resolution of the instrument impose further restrictions on the nano-particle's mass. Larger nano-particle masses lead to a narrower width of the wave packet, escalating the requirements for current stability and instrument spatial resolution to observe spatial interference fringes, see Appendices \ref{current_fluctuation} and \ref{Spreading_of_WP}.
 
In this scheme, the separation and closing of the wave packet trajectories can be achieved using a modest current density ${\cal O}(1)~{\rm A/\mu m^{2}}$. This capability allows for the effective recombination of wave packet trajectories using the SG apparatus, leading to the eventual restoration of spin coherence \cite{Marshman:2021wyk, zhou2022}.  However, a detailed exploration of these latter aspects will be addressed independently. Additionally, several considerations merit attention, including the coherence of spin in the presence of the NV center \cite{Japha:2022xyg,Japha:2022phw}, and the excitations of the phonons~\cite{Henkel:2021wmj}. Such considerations are left for future study. Since the diamagnetic enhancement does not necessitate spin-based manipulation, there may be a simpler candidate for creating the initial small spatial splitting without grappling with rotation-related challenges \cite{Japha:2022xyg,Japha:2022phw}. Nevertheless, it is likely that comparable constraints will emerge to ensure the feasibility of coherent interference.

{\bf Acknowledgements}
	R. Z. is supported by China Scholarship Council (CSC) fellowship. R. J. M. is supported by the Australian Research Council (ARC) under the Centre of Excellence for Quantum Computation and Communication Technology (CE170100012). S. B. would like to acknowledge EPSRC grants (EP/N031105/1, EP/S000267/1 and EP/X009467/1) and grant ST/W006227/1. 

\bibliography{references} 
\begin{appendices}
	
\section{Initial separation}\label{initial_condition}
The process of enlarging the spatial superposition size from 1 ${\rm \mu m}$ by a factor of 100 to 1000 ${\rm \mu m}$ is, notably, mass-independent. However, achieving the initial separation appears to be unavoidably mass-dependent, as explained in the main text in the conclusion and discussion section. To obtain spatial superposition states for diamagnetic nano-particles, a potentially effective approach involves leveraging the interaction of the NV spins embedded in them with the magnetic field. The corresponding Hamiltonian, encompassing the NV spin-magnetic field interaction, is expressed as follows \cite{wan2016free,Pedernales_2020,Marshman:2021wyk,Wood:2022htp}:
\begin{equation}
	\hat{H}=\frac{\hat{\boldsymbol{p}}^{2}}{2 m}+\hbar D \hat{S}_{z'}^{2}-\hat{\boldsymbol{\mu}}\cdot \vb{B} - \frac{\chi_{\rho} m}{2 \mu_{0}} \vb{B}^{2} + mg\hat{\boldsymbol{z}}.
\end{equation}	
Here, the second term represents the zero-field splitting of the NV center with 
$D=(2\pi)\times2.8~{\rm GHz}$, $\hbar$ is the reduced Planck constant, and $\hat{S}_{z'}$ is the spin component operator aligned with the NV axis. The third term denotes the interaction energy of the NV electron spin magnetic moment with the magnetic field $\vb{B}$. The spin magnetic moment operator 
$\hat{\boldsymbol{\mu}}=-g_{s}\mu_{B}\hat{\boldsymbol{S}}$, where $g_{s}\approx 2$ is the Land\`{e} g-factor, $\mu_{B}=e \hbar/2 m_{e}$ is the Bohr magneton, and $\hat{\boldsymbol{S}}$ is the NV spin operator with eigenstates denoted by \{$\ket{+1}, \ket{-1}, \ket{0}$\}.

Assuming the initial superposition state as $(\ket{+1}+\ket{-1})/\sqrt{2}$, initially, the only factor separating the two wave packets is the NV spin-magnetic field interaction term $\hat{\boldsymbol{\mu}}\cdot \vb{B}$. The corresponding acceleration is given by:
\begin{align}\label{acceleration_for_spin}
	a_{\pm} = \mp \frac{g_{s}\mu_{B}}{m}\nabla \text{B},
\end{align}
where the subscript $``\pm"$ corresponds to the $\ket{+1}$ and $\ket{-1}$ states, respectively. From Eq.(\ref{acceleration_for_spin}), it is evident that the acceleration is inversely proportional to the mass, implying that the initial spatial separation produced using the spin-magnetic field interaction is also inversely proportional to the mass. Assuming a linear magnetic field approximation 
$B = \eta x$ ($\eta$ is the magnetic field gradient), the initial spatial separation between wave packets with respect to mass is \cite{Pedernales_2020}:
\begin{align}\label{initial_separation}
	D_{ini} = \frac{2g_{s}\mu_{B}\mu_{0}}{-\chi_{\rho}}\frac{1}{m\eta}\qty(\cos(\omega t) - 1),
\end{align}
where
\begin{align}\label{angular_frequency}
	\omega = \sqrt{\frac{-\chi_{\rho} }{\mu_0}}\eta 
\end{align}
is the angular frequency. Notably, the initial separation in Eq.(\ref{angular_frequency}) is not only inversely proportional to the mass but also to the magnetic field gradient. This implies that arbitrarily large spatial superposition sizes can be achieved as long as the magnetic field gradient is sufficiently small. However, considering the periodic motion of diamagnetic nanoparticles in a linear magnetic field, and accounting for decoherence effects, for a $10^{-15}$ kg nanoparticle with a spin coherence time limited to 0.5 s, the required magnetic field gradient at this point is approximately 45 T/m, corresponding to a maximum spatial separation of 0.2 ${\rm \mu m}$. Even with such a small initial separation, the current density needed to enhance the spatial superposition size using wires is within reasonable limits of about 10
${\rm A/\mu m^{2}}$ \cite{Yao2000,Wei2001,Murali2009}.
	
After obtaining the initial separation using the NV spin, two approaches can be employed to integrate it with the enhanced spatial superposition size scheme. The first approach involves retaining the NV spin and placing the nano-particle into the magnetic field generated by the wire. However, in this case, the direction of the spin-magnetic field interaction force on the wave packets with spin eigenstates 
$\ket{-1}$ and $\ket{+1}$ are in the direction of and against the magnetic field gradient, respectively. This leads to asymmetrical trajectories for the two wave packets, complicating the scattering process. To avoid this complexity, an alternative method is employed. Initially, a $\pi/2$ pulse is applied to transform the $(\ket{+1}+\ket{-1})/\sqrt{2}$ state into the $\ket{0}$ state \cite{TaylorEtAl2008,levine2019}. This transformation renders the interaction of the NV spin with the magnetic field negligible, reducing the Hamiltonian to the form in Eq.(\ref{Hamitonian}).
	
\section{Limitation on current fluctuation for generating spatial interference}\label{current_fluctuation}

According to Equation (\ref{Acceleration}), fluctuations in the electrical current can introduce classical uncertainties in both the position and momentum of an object. These fluctuations have the potential to impact various quantum phenomena, including spatial interference \cite{MargalitEtAl2019}, momentum interference \cite{MachlufEtAl2013}, and spin coherence \cite{Folman2018} between wave packets. In our specific scenario, a nano-particle undergoes elastic collisions with a stationary wire. Importantly, fluctuations in the current affect only the direction of the nano-particle's velocity, not its magnitude. Consequently, our primary focus here is on the perturbation of the nano-particle's spatial position. For the sake of simplicity, we assume that the classical position and momentum of the nano-particle at its initial state are error-free.

\begin{figure}
	\includegraphics[width=0.7\linewidth]{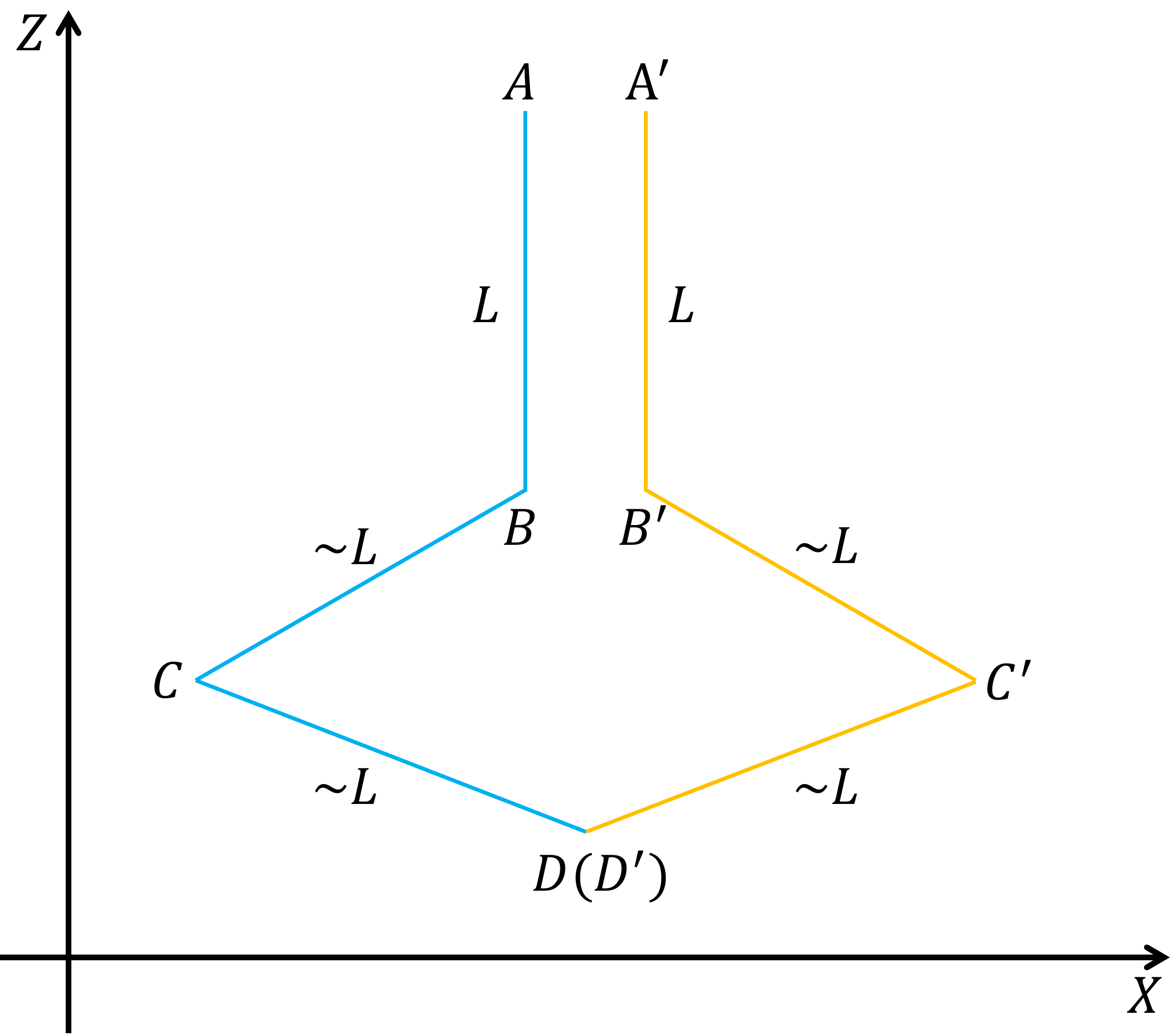}
    \caption{ Illustration of wave packet trajectories. The diagram depicts the trajectories of two wave packets with blue and orange line segments. Each trajectory consists of three segments, approximately of length $L$. Points $\text{A}(\text{A}^{'})$ and $\text{D}(\text{D}^{'})$ correspond to the initial and final positions of the wave packet trajectories. Points $\text{B}(\text{B}^{'})$ and $\text{C}(\text{C}^{'})$ indicate the positions where the wave packets scatter from the wire. It is important to note that the angles between the line segments in the figure do not accurately represent the true scattering angles.} \label{DraftTrajectory}
\end{figure}

In the subsequent analysis, we quantitatively assess how the trajectory of the nano-particle deviates following its interactions with the wire, both after the initial scattering and the subsequent scattering, while considering a small current fluctuation, denoted as $\Delta I$. To streamline our calculations, we model the nano-particle's trajectory as comprising discrete line segments, each with a fixed length $L$, as depicted in Fig.\ref{DraftTrajectory}. Given the symmetry in the trajectories of the left and right wave packets, we choose to illustrate the left trajectory (represented by the blue line segment) for clarity.

\begin{figure}
	\includegraphics[width=0.8\linewidth]{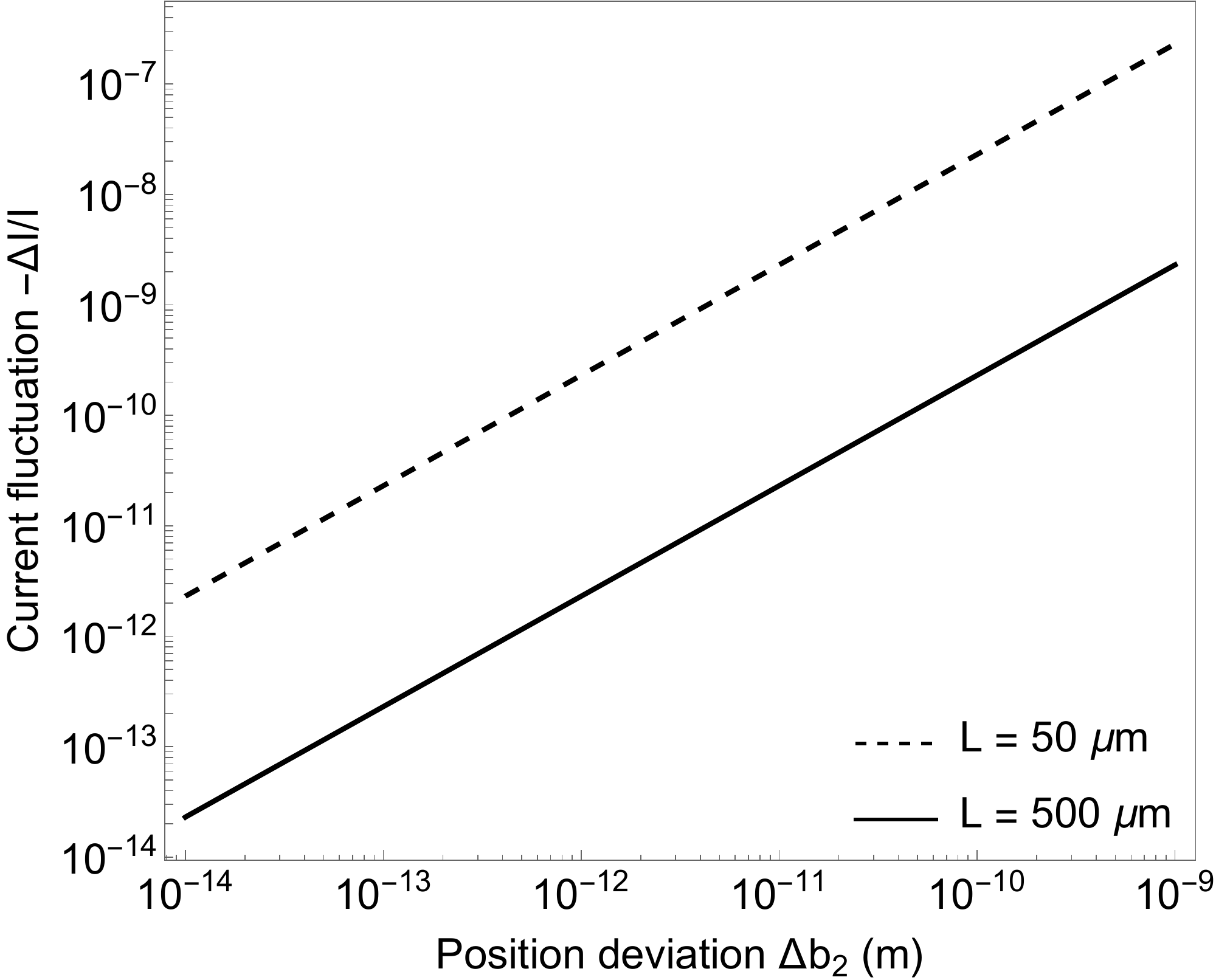}
	\caption{Current fluctuation versus position deviation of the wave packet (at point $\text{D}$). The black dashed line corresponds to $L=50\,{\rm \mu m}$ and the black solid line corresponds to $L=500\,{\rm \mu m}$. $L$ is defined in Fig.\ref{DraftTrajectory}. We set $b=5\times 10^{-7}$\,{\rm m}.} \label{CurrentFluctuation}
\end{figure}

Utilizing the scattering angle expression from Eq.(\ref{ScatteringAngle}) to differentiate with respect to both current and impact parameter, we obtain
\begin{align}\label{ChangeInSA1}
	\Delta \theta_{s}&=\beta\qty(\frac{\Delta I}{I}-\frac{\Delta b}{b}),\nonumber\\
	&=\Delta \theta_{sI}+\Delta \theta_{sb},
\end{align}
where $\Delta \theta_{sI}=\beta\Delta I/I$ represents the change in scattering angle due to fluctuations in current, and $\Delta \theta_{sb}=-\beta\Delta b/b$ represents the change in scattering angle due to fluctuations in the impact parameter. Here, $\beta$ is a coefficient associated with the current $I$, incident velocity $v_{in}$, and impact parameter $b$ with the expression
\begin{equation}
	\beta=\frac{k-1}{k^{\frac{3}{2}}}\pi.
\end{equation}
The expression for the parameter $k$ is provided in Eq.(\ref{Frequency}). Specifically, for the first scattering event, where $\theta_{s1}\approx \pi/2$, we have $\beta_{1}=3\pi/8$. For the second scattering, where $\theta_{s2}\approx 3\pi/4$, we have $\beta_{2}=15\pi/64$. According to Eq.(\ref{ChangeInSA1}), we can express the classical trajectory deviation of the wave packet as it reaches point C as
\begin{equation}\label{IPUncertainty1}
	\Delta b_{1}\approx \Delta \theta_{sI}(\text{B})\times L=\beta_{1}\frac{\Delta I}{I}L.
\end{equation}
where the term 'B' in brackets denotes fluctuations in the scattering angle at point B, similarly thereafter. By combining Eq.(\ref{ChangeInSA1}) with Eq.(\ref{IPUncertainty1}), we can determine the uncertainty in the scattering angle at the second scattering event as follows
\begin{align}\label{ChangeInSA2}
	\Delta \theta_{s}(\text{C})&=\Delta \theta_{sI}(\text{C})+\Delta \theta_{sb}(\text{C}),\nonumber\\
	&=\beta_{2}\frac{\Delta I}{I}-\beta_{1}\beta_{2}\frac{\Delta I}{I}\frac{L}{b}.
\end{align}
Since $\beta_{1}$ is approximately equal to 1, when the condition $L \gg b$ holds, it becomes evident that $\Delta \theta_{sI}(\text{C}) \ll \Delta \theta_{sb}(\text{C})$. Therefore, for the second scattering event, we focus solely on $\Delta\theta_{sb}(\text{C})$. When the wave packet reaches point D, the deviation in the classical trajectory can be expressed as
\begin{equation}\label{IPUncertainty2}
	\Delta b_{2}\approx \Delta \theta_{s}(C)\times L = -\beta_{1}\beta_{2}\frac{\Delta I}{I}\frac{L^{2}}{b}.
\end{equation}
Rewrite Eq.(\ref{IPUncertainty2}) as
\begin{equation}\label{LimitationofCurrent}
	-\frac{\Delta I}{I}=\frac{b}{\beta_{1}\beta_{2}L^{2}}\Delta b_{2}.
\end{equation}
To ensure spatial interference at a specific point $\text{D}(\text{D}^{'})$, it is imperative that $\Delta b_{2}$ remains smaller than the corresponding wave packet width. This requirement imposes constraints on the fluctuations in the current. Based on Eq.(\ref{LimitationofCurrent}), we have depicted the relationship between current fluctuation $\Delta I/I$ and position deviation $\Delta b_2$ in Fig.\ref{CurrentFluctuation}. Notably, when the final requirement for position deviation remains constant, reducing the superposition size $L$ by an order of magnitude alleviates the demand for current stability by two orders of magnitude.

To illustrate this concept with a practical example, let us consider a silica microsphere with a mass of $10^{-15}\,{\rm kg}$ trapped by a magnetic field oscillating at a frequency of 100 Hz \cite{slezakSilica2018}. The initial width of the wave packet can be calculated as
\begin{align}\label{WavepacketWidth}
	\delta x&=\sqrt{\frac{\hbar}{2m\omega}}\approx 2\times 10^{-11}\,{\rm m},\nonumber\\
	\delta p&=\sqrt{\frac{\hbar m\omega}{2}}\approx 2.3\times 10^{-24}\, {\rm kg\,m/s}.
\end{align}
For spatial interference to occur, $\Delta b_{2}$ must be smaller than $\delta x$. Combining Eq.(\ref{LimitationofCurrent}) and Eq.(\ref{WavepacketWidth}), we derive the following limits on current fluctuations
\begin{align}\label{LimitationofCurrent2}
&\abs{\frac{\Delta I}{I}}
\lesssim\begin{cases}
	5\times 10^{-11}\quad \text{for}\,L=500\,{\rm \mu m},\\
	\,\\
	5\times 10^{-9}\quad\,\,\, \text{for}\,L=50\,{\rm \mu m}.
\end{cases}
\end{align}
However, it is essential to acknowledge that as the wave packet evolves with time, the actual limits on current fluctuations may be less stringent than those presented in Eq.(\ref{LimitationofCurrent2}).

\section{The spreading of wave packet}\label{Spreading_of_WP}

Here, we give the estimation of the change in the width of the wave packet in both the free evolution case and after it has been scattered by the wire. The true width of the wave packet should lie between these two cases. This discussion focuses on the one-dimensional scenario to simplify the analysis. Assuming that an initial state represented by a Gaussian wave packet, the probability density for free evolution is given by \cite{Buoninfante:2017rbw}:
\begin{align}
	\rho(x, t)=\left(\frac{1}{2 \pi}\right)^{\frac{1}{2}} \sqrt{\frac{\delta_{x}^{2}}{\delta_{x}^{4} + \alpha^{2} t^{2}}} e^{- \frac{1}{2} \frac{\delta_{x}^{2}}{\delta_{x}^{4} + \alpha^{2} t^{2}} (x-x_{0}-v_{g}t)^{2}}.
\end{align}
Here, $x_{0}$ represents the initial classical position, and $v_{g}$ denotes the group velocity of the wave packet. The probability density follows a Gaussian distribution, and the width of the Gaussian function is determined by:
\begin{equation}\label{evolution_of_wave_packet}
	\delta_{x}(t)=\sqrt{\delta_{x}^{2} + \frac{\alpha^{2} t^{2}}{\delta_{x}^{2}}},
\end{equation}
where $\alpha=\hbar/2 m$ is defined. For a wave packet with a mass of $m=10^{-15}\,\mathrm{kg}$ and an initial width of $2\times 10^{-11}\,{\rm m}$ (as given in Eq.(\ref{WavepacketWidth})), the width evolves after a time $t=0.02\,{\rm s}$, yielding:
\begin{equation}
	\delta_{x}(0.02\,\text{s})\approx 5.6\times 10^{-11}\,{\rm m}.
\end{equation}
This result indicates that, after 0.02 s of free evolution, the width of the wave packet remains approximately within the same order of magnitude as the initial width.

To give a rough order of magnitude estimate of the change in the wave packet width following scattering, we invoke the uncertainty principle, expressed as:
\begin{equation}\label{uncertainty_principle}
	\delta x\ge\frac{\hbar}{2\delta p}.
\end{equation}
Setting the scattering time scale as $\delta t = b/v_{in}$, we establish a lower bound on the width of the wave packet post-scattering by placing the classical position of the wave packet at a distance of 1 $\mathrm{\mu m}$ from the wire. This proximity represents the point of the strongest interaction between the wave packet and the magnetic field in our scheme. The change in velocity of the wave packet during scattering is given by:
\begin{align}\label{change_in_velocity}
	\delta v &= a_{dia}\delta t,\nonumber\\
	&\approx 3.6\times 10^{-2}\,\mathrm{m/s}.
\end{align}
Considering a mass of $10^{-15}$ kg, combining Eq.(\ref{uncertainty_principle}) and (\ref{change_in_velocity}), we derive a lower bound on the width of the wave packet after scattering:
\begin{align}\label{limit_on_width_of_WP}
	\delta x &\ge \frac{\hbar}{2m \delta v},\nonumber\\
	&\approx 1.5\times 10^{-18}\,\text{m}.
\end{align}
The implication of Eq.(\ref{limit_on_width_of_WP}) is that the wave packet exhibits more particle-like characteristics after scattering. 
This suggests that, for the generation and detection of spatial interferences, increased demands are placed on the stability of the current (refer to Eq.(\ref{LimitationofCurrent})) and the spatial resolution of the measurement instrument. Two potential mitigation strategies are identified. The first involves reducing the mass of the nano-particle, for instance, to $10^{-22}$ kg, resulting in a position uncertainty $\delta x \ge 1.5\times 10^{-11}\,\text{m}$. The second method is to apply a magnetic field to trap the nano-particle so that the spatial width of the wave packet spreads out rapidly,  as described in Eq.(\ref{WavepacketWidth}).

\end{appendices}

\end{document}